\documentclass[twoside,twocolumn,9pt]{article}
\usepackage{extsizes}
\usepackage[super,sort&compress,comma]{natbib} 
\usepackage[version=3]{mhchem}
\usepackage[left=1.5cm, right=1.5cm, top=1.785cm, bottom=2.0cm]{geometry}
\usepackage{balance}
\usepackage{mathptmx}
\usepackage{sectsty}
\usepackage{graphicx} 
\usepackage{lastpage}
\usepackage[format=plain,justification=justified,singlelinecheck=false,font={stretch=1.125,small,sf},labelfont=bf,labelsep=space]{caption}
\usepackage{float}
\usepackage{fancyhdr}
\usepackage{fnpos}
\usepackage{siunitx}
\usepackage{mhchem}
\usepackage[english]{babel}
\addto{\captionsenglish}{%
  
}
\usepackage{array}
\usepackage{droidsans}
\usepackage{charter}
\usepackage[T1]{fontenc}
\usepackage[usenames,dvipsnames]{xcolor}
\usepackage{setspace}
\usepackage[compact]{titlesec}
\usepackage{hyperref}
\usepackage{url}
\usepackage{color}
\usepackage{epstopdf}%This line makes .eps figures into .pdf - please comment out if not required.

\definecolor{cream}{RGB}{222,217,201}

\begin{document}

\pagestyle{fancy}
\thispagestyle{plain}
\fancypagestyle{plain}{
%%%HEADER%%%
\renewcommand{\headrulewidth}{0pt}
}
%%%END OF HEADER%%%

%%%PAGE SETUP - Please do not change any commands within this section%%%
\makeFNbottom
\makeatletter
\renewcommand\LARGE{\@setfontsize\LARGE{15pt}{17}}
\renewcommand\Large{\@setfontsize\Large{12pt}{14}}
\renewcommand\large{\@setfontsize\large{10pt}{12}}
\renewcommand\footnotesize{\@setfontsize\footnotesize{7pt}{10}}
\makeatother

\renewcommand{\thefootnote}{\fnsymbol{footnote}}
\renewcommand\footnoterule{\vspace*{1pt}% 
\color{cream}\hrule width 3.5in height 0.4pt \color{black}\vspace*{5pt}} 
\setcounter{secnumdepth}{5}

\makeatletter 
\renewcommand\@biblabel[1]{#1}            
\renewcommand\@makefntext[1]% 
{\noindent\makebox[0pt][r]{\@thefnmark\,}#1}
\makeatother 
\renewcommand{\figurename}{\small{Fig.}~}
\sectionfont{\sffamily\Large}
\subsectionfont{\normalsize}
\subsubsectionfont{\bf}
\setstretch{1.125} %In particular, please do not alter this line.
\setlength{\skip\footins}{0.8cm}
\setlength{\footnotesep}{0.25cm}
\setlength{\jot}{10pt}
\titlespacing*{\section}{0pt}{4pt}{4pt}
\titlespacing*{\subsection}{0pt}{15pt}{1pt}
%%%END OF PAGE SETUP%%%

%%%FOOTER%%%
\fancyfoot{}
\fancyfoot[LO,RE]{\vspace{-7.1pt}\includegraphics[height=9pt]{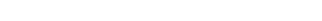}}
\fancyfoot[CO]{\vspace{-7.1pt}\hspace{13.2cm}\includegraphics{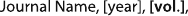}}
\fancyfoot[CE]{\vspace{-7.2pt}\hspace{-14.2cm}\includegraphics{head_foot/RF}}
\fancyfoot[RO]{\footnotesize{\sffamily{1--\pageref{LastPage} ~\textbar  \hspace{2pt}\thepage}}}
\fancyfoot[LE]{\footnotesize{\sffamily{\thepage~\textbar\hspace{3.45cm} 1--\pageref{LastPage}}}}
\fancyhead{}
\renewcommand{\headrulewidth}{0pt} 
\renewcommand{\footrulewidth}{0pt}
\setlength{\arrayrulewidth}{1pt}
\setlength{\columnsep}{6.5mm}
\setlength\bibsep{1pt}
%%%END OF FOOTER%%%

%%%FIGURE SETUP - please do not change any commands within this section%%%
\makeatletter 
\newlength{\figrulesep} 
\setlength{\figrulesep}{0.5\textfloatsep} 

\newcommand{\topfigrule}{\vspace*{-1pt}% 
\noindent{\color{cream}\rule[-\figrulesep]{\columnwidth}{1.5pt}} }

\newcommand{\botfigrule}{\vspace*{-2pt}% 
\noindent{\color{cream}\rule[\figrulesep]{\columnwidth}{1.5pt}} }

\newcommand{\dblfigrule}{\vspace*{-1pt}% 
\noindent{\color{cream}\rule[-\figrulesep]{\textwidth}{1.5pt}} }

\makeatother
%%%END OF FIGURE SETUP%%%

%%%TITLE, AUTHORS AND ABSTRACT%%%
\twocolumn[
  \begin{@twocolumnfalse}
{\includegraphics[height=30pt]{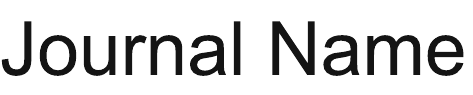}\hfill\raisebox{0pt}[0pt][0pt]{\includegraphics[height=55pt]{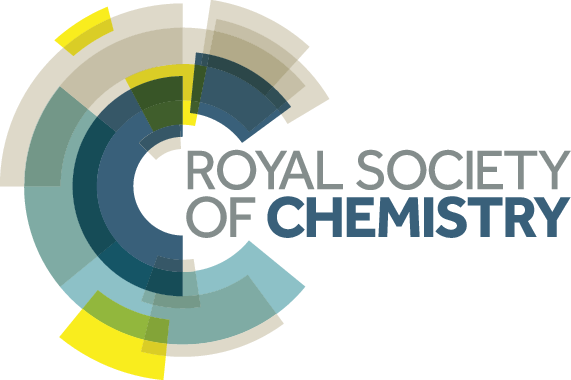}}\\[1ex]
\includegraphics[width=18.5cm]{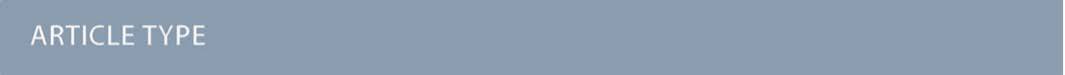}}\par
\vspace{1em}
\sffamily
\begin{tabular}{m{4.5cm} p{13.5cm} }

\includegraphics{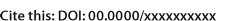} & \noindent\LARGE{\textbf{Universal neural network potentials as descriptors:  Towards scalable chemical property prediction using quantum and classical computers}} \\%Article title goes here instead of the text "This is the title"
\vspace{0.3cm} & \vspace{0.3cm} \\

 & \noindent\large{Tomoya Shiota,$^{\ast}$\textit{$^{a, b}$} Kenji Ishihara,\textit{$^{b}$} and Wataru Mizukami\textit{$^{{\ast}a, b}$}} \\%Author names go here instead of "Full name", etc.

\includegraphics{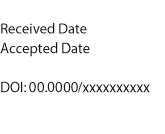} & \noindent\normalsize{Accurate prediction of diverse chemical properties is crucial for advancing molecular design and materials discovery. Here we present a versatile approach that uses the intermediate information of a universal neural network potential as a general-purpose descriptor for chemical property prediction. Our method is based on the insight that by training a sophisticated neural network architecture for universal force fields, it learns transferable representations of atomic environments. We show that transfer learning with graph neural network potentials such as M3GNet and MACE achieves accuracy comparable to state-of-the-art methods for predicting the NMR chemical shifts of using quantum machine learning as well as a standard classical regression model, despite the compactness of its descriptors. In particular, the MACE descriptor demonstrates the highest accuracy to date on the ${^{13}}$C NMR chemical shift benchmarks for drug molecules. This work provides an efficient way to accurately predict properties, potentially accelerating the discovery of new molecules and materials.} \\%The abstrast goes here instead of the text "The abstract should be..."

\end{tabular}

 \end{@twocolumnfalse} \vspace{0.6cm}

  ]
%%%END OF TITLE, AUTHORS AND ABSTRACT%%%

%%%FONT SETUP - please do not change any commands within this section
\renewcommand*\rmdefault{bch}\normalfont\upshape
\rmfamily
\section*{}
\vspace{-1cm}

%%%FOOTNOTES%%%

\footnotetext{\textit{$^{a}$Graduate School of Engineering Science, Osaka University, 1-3 Machikaneyama, Toyonaka, Osaka 560-8531, Japan; E-mail: shiota.tomoya.ss@gmail.com; mizukami.wataru.qiqb@osaka-u.ac.jp}}
\footnotetext{\textit{$^{b}$Center for Quantum Information and Quantum Biology, Osaka University, 1-2 Machikaneyama, Toyonaka 560-8531, Japan}}
\footnotetext{\textit{$^{\ast}$Corresponding auther}}
%Please use \dag to cite the ESI in the main text of the article.
%If you article does not have ESI please remove the the \dag symbol from the title and the footnotetext below.
%additional addresses can be cited as above using the lower-case letters, c, d, e... If all authors are from the same address, no letter is required

%%%END OF FOOTNOTES%%%

%%%MAIN TEXT%%%%

\section{\label{sec:level1}Introduction}

As evidenced by the enumeration of 166.4B possible organic molecules containing up to 17 heavy elements, such as C, N, O, S, and halogens (excluding hydrogen), the expansion of the chemical space is astronomical with the increase in types and numbers of elements.\cite{Reymond_2010, Ruddigkeit_2012} This vast landscape has given rise to multidisciplinary approaches to combining experimental and computational chemistry for the discovery of new chemical substances and materials in a wide range of fields, including material, catalysis, and drug design.\cite{Reymond_2010, Ruddigkeit_2012, G_mez_Bombarelli_2016, Curtarolo_2013, Nie_2023, Oganov_2019} Although quantum chemistry and first-principles calculations offer accurate descriptions of chemical substances, their high computational demands make an exhaustive exploration of the chemical space impractical.\cite{pierens20141h, Hartman_2016, Kleine_B_ning_2023, Chen_2022, Lodewyk_2011, Lauro_2020, Oganov_2019} However, machine- and deep-learning techniques are overcoming these limitations to enable a more extensive exploration.\cite{hansen2015machine, Deringer_2021, Faber_2017, Sajjan_2022, Keith_2021, Chen_2022, Wan_2023, Wu_2018, Reiser_2022, Bart_k_2017, Curtarolo_2013, Kocer_2022, langer2022representations, Liu_2021, Merchant_2023,Gao_2020, Macalino_2015, Oganov_2019} 

With machine learning, physics-inspired descriptors that characterize the chemical space have been developed and serve as the cornerstone for building efficient and highly accurate models.\cite{HIMANEN2020106949, Gupta_2021, Gerrard_2020, Willatt_2019, Bart_k_2013, Faber_2018, Christensen_2020, Kabylda_2023, Musil_2021, Bart_k_2010, Bart_k_2017, Parsaeifard_2021, langer2022representations, Khan_2023, Rupp_2015} 
Smooth overlap of atomic positions (SOAP)\cite{HIMANEN2020106949, Deringer_2021, Cordova_2022, Willatt_2019, Bart_k_2013, Wan_2023, Bart_k_2017}, Faber–Christensen–Huang–Lilienfeld (FCHL)\cite{Gupta_2021, Gerrard_2020, Faber_2018, Christensen_2020, Parsaeifard_2021}, and similar descriptors offer atom-level descriptions within molecular or material environments based on physical insights and are effective in regressing chemical quantities, such as interatomic potentials (IAP) and nuclear magnetic resonance (NMR) chemical shifts.\cite{HIMANEN2020106949, Deringer_2021, Cordova_2022, Kohlhoff_2009, Gupta_2021, Gerrard_2020, Faber_2017, Willatt_2019, Bart_k_2013, Faber_2018, Christensen_2020, Kabylda_2023, Musil_2021, Lodewyk_2011, Bart_k_2010, Bart_k_2017, Parsaeifard_2021, langer2022representations, Liu_2021, Tsitsvero_2023} Notably, IAPs built using descriptors and Gaussian process regression (GPR) \cite{Deringer_2021} have been termed Gaussian approximation potentials (GAP) and have found success in the exploration of the chemical space of molecules and materials.\cite{Deringer_2021, Bart_k_2010, Bart_k_2017} Both kernel ridge regression (KRR) and GPR have been employed to improve the accuracy of NMR chemical shift prediction \cite{Paruzzo_2018, Cordova_2022, Kohlhoff_2009, Gupta_2021, Gerrard_2020, Tsitsvero_2023}. However, the dimensionality of the descriptors becomes a barrier to generalization and high accuracy as the molecular or material composition becomes more diverse owing to the addition of different types of elements.\cite{Kabylda_2023, Wu_2018, Fung_2021, Khan_2023}

Recently, deep-learning models based on graph neural networks (GNNs) have been proposed to describe chemical spaces using graph representations.\cite{Guan_2021, Liu_2019, Han_2021, Kwon_2020, Jonas_2019, Kleine_B_ning_2023, Keith_2021, Wan_2023, schütt2017schnet, Chen_2019, Wu_2018, gasteiger2021gemnet, liao2023equiformer,liao2024equiformerv2, Reiser_2022, Fung_2021, xu2019powerful, pmlr-v162-stark22a, Jiang_2021, Liu_2021, Merchant_2023, Kang_2020, takamoto2022towards, Chen_2022, musaelian2023learning, Batatia2022mace, kovacs2023mace, batatia2023foundation} In most GNN-based IAPs, atoms within a molecular or material environment are represented as nodes, and their local connectivity as edges in a graph. The graph is then convolved to embed atom-specific information within each node, and further processed using multilayer perceptrons (MLP) to predict target observables. In molecular and materials simulation and modeling, the consideration of symmetry is extremely important. It is desirable for GNNs to be invariant or equivariant to symmetry operations such as translation, rotation, and reflection for the models to make physically meaningful predictions. GNNs that possess these properties are referred to as invariant GNNs or equivariant GNNs. The universal GNN-based IAPs proposed thus far have been designed to satisfy these symmetries. Recently, E(3) or SE(3) equivariant GNN-based IAPs (e.g., Allegro\cite{musaelian2023learning}, GNoME\cite{merchant2023scaling}, MACE\cite{Batatia2022mace, kovacs2023mace, batatia2023foundation}) have demonstrated superior performance compared to E(3) invariant GNN-based IAPs (e.g., MEGNet\cite{Chen_2019}, M3GNet\cite{Chen_2022}).\cite{batzner20223, riebesell2024matbench}

Similarly, GNN-based models have been developed to predict NMR chemical shifts.\cite{Guan_2021, Liu_2019, Kwon_2020, Jonas_2019, Han_2022, Kang_2020} DFT-level calculations of NMR chemical shifts for $^1$H and $^{13}$C have demonstrated the ability to predict within a target accuracy range of 1-2\% relative to the possible ranges of approximately 10 ppm and 200 ppm, respectively\cite{grimblat2015beyond, semenov2020dft}. Therefore, the uncertainty in machine learning models using DFT-level datasets is this level of precision, with the target accuracy of 0.2 ppm for $^1$H and 2 ppm for $^{13}$C.\cite{Gerrard_2020} For example, Yanfei Guan et al. achieved the target accuracy of \SI{0.16}{ppm} for $^1$H and \SI{1.26}{ppm} for $^{13}$C by training the SchNet architecture \cite{schütt2017schnet} on molecular NMR chemical shifts (CASCADE)\cite{Guan_2021}.

However, the scalability remains an issue due to the increasing optimization costs of GNN and MLP parameters when the size of datasets increase. Han et al. addressed this issue by constraining the nodes in a GNN to heavy elements only, thereby rendering the construction of scalable GNN-based NMR chemical shift models feasible while achieving a state-of-the-art prediction accuracy comparable to that of CASCADE.\cite{Han_2022} Furthermore, NMR chemical shifts of various nuclei beyond hydrogen and carbon have become crucial for understanding systems involving a wide range of elements, such as proteins and solids.\cite{schaefer1986mechanisms, fukaya2004dft, chen201319f, yu2013new, gerrard2022prediction, krivdin202317o, matsuzaki2024origin} Consequently, efforts are being made to develop machine learning models for NMR chemical shifts of nuclei such as $^{15}$N, $^{17}$O, and $^{19}$F.  \cite{gerrard2022prediction, krivdin202317o, chen201319f, yu2013new} These elements exhibit wide chemical shift ranges, with about 600, 2500, 500 ppm for $^{15}$N, $^{17}$O, and $^{19}$F, respectively. The target accuracy for these nuclei is set at 25 ppm for $^{15}$N and 5 ppm for $^{19}$F as well as $^1$H and $^{13}$C.\cite{schaefer1986mechanisms, fukaya2004dft, chen201319f, yu2013new, gerrard2022prediction, krivdin202317o, matsuzaki2024origin}

Notably, both descriptor-based and GNN-based methods face challenges. The former faces increased learning costs as the composition becomes more complex, and the latter faces increasing parameter optimization costs with larger training datasets. To address these issues simultaneously, we focused on the potential utility of the outputs from pre-trained GNN-based IAPs as descriptors. We considered these outputs GNN transfer learning (GNN-TL) descriptors and built machine-learning models for predicting chemical properties. Note that there are existing studies attempting to apply pre-trained GNN potentials to other tasks, particularly to generative modeling.~\cite{xie2021crystal, wu2022diffusion, zaidi2022pre, jia2024derivative} 

The remainder of this paper is organized as follows. Section ~\ref{sec:Method} details the GNN-TL descriptor and the kernel method, implemented on both classical and quantum computers, for predicting NMR chemical shifts of $^1$H, $^{13}$C, $^{15}$N, $^{17}$O, and $^{19}$F. Section ~\ref{sec:Results}  presents the performance of our developed machine learning models. Section ~\ref{sec:Discussion} discusses the benefits and applications of the GNN-TL descriptor. Finally, Section ~\ref{sec:Conclusion} concludes the paper.

\begin{figure*}[t]
\centering
\includegraphics{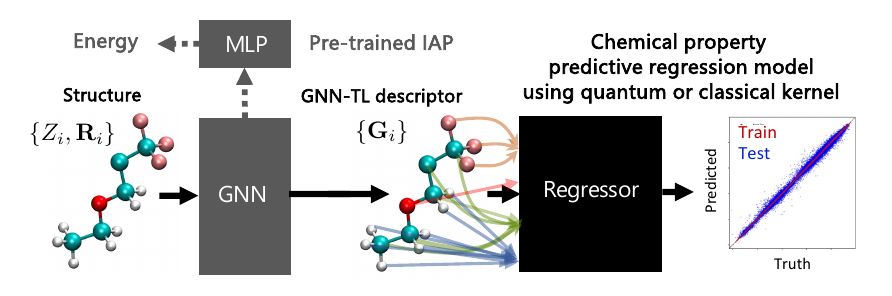}
\caption{\label{fig:1} Schematic diagram of our proposed graph neural network transfer learning for predicting chemical properties. The black arrows depict the flow of our transfer learning process. The gray area is a pre-trained IAP (NNP) designed for predicting the energy of the system and composed of a GNN and an MLP. The initial step in our learning procedure involves obtaining the pre-trained GNN block output a set of vectors, $\left\{\mathbf{G}_i\right\}$, using the atomic coordinates of a molecule with $N$ atoms, $\left\{Z_i, \mathbf{R}_i\right\}$ as input. Subsequently, we construct a regression model to predict the chemical properties $e. g.$ NMR shielding constants, using this GNN output $\left\{\mathbf{G}_i\right\}$ as a descriptor.}
\end{figure*}

\section{\label{sec:Method}Method: Transfer Learning Using Pre-trained Graph Neural Network}

In this section, we discuss the transfer learning of a pre-trained GNN-based IAP. This approach integrates the outputs from the GNN layer of the IAP as shown in Fig. 1. The architecture of a GNN-based IAP can be broadly segmented into a GNN layer and an MLP layer (gray area of Fig. 1). For the E(3) invariant GNN-based IAP, we opted for two backbones: a MEGNet pre-trained on the QM9 dataset\cite{Ramakrishnan_2014} and a M3GNet trained on the MPF.2021.2.8 dataset, which encompasses compounds covering all 89 elements from the Materials Project.\cite{Chen_2022} The parameters of the GNN layer in the M3GNet IAP were optimized to predict system energy, forces, and stress tensors. Additionally, we incorporated the E(3) equivariant GNN-based IAPs, namely MACE\cite{Batatia2022mace, batatia2023foundation, kovacs2023mace}, into our study. We employed two types of pre-trained MACE IAPs: one trained on a larger dataset named MPtrj\cite{deng2023chgnet} from Materials Project, referred to as the MACE-MP0 model\cite{batatia2023foundation}, and another trained on an organic molecule dataset covering 10 types of elements including SPICE\cite{eastman2023spice} and QMug\cite{isert2022qmugs}, termed the MACE-OFF23 model\cite{kovacs2023mace}. Each model has variations in parameter size, and for this study, we utilized the "small" and "large" versions.\cite{batatia2023foundation, kovacs2023mace} 

When fed with the atomic coordinates of a molecule with $N$ atoms, denoted by $\left\{Z_i, \mathbf{R}_i\right\}$, where $Z_i$ represents the atomic number indicating the type of each atom, and $\mathbf{R}_i$ is the three-dimensional position vector of the $i$th atom, the GNN layer generates a set of vectors, {$\left\{\mathbf{G}_i\right\}$}, which mirrors the environment of the $i$th atom in the molecule. This is referred to as the GNN-TL descriptor. The GNN layer for both MEGNet and M3GNet outputs GNN-TL descriptors with dimensions of 32 and 64 per atom, respectively.  On the other hand, MACE is a GNN architecture that predicts energy in the form of atomic cluster expansion. As in Ref.\cite{elijovsius2024zero}, only the output of the 1st layer of the GNN layer, corresponding to the one-body term of the many-body expansion, is used as the GNN-TL descriptor. The dimensions of this GNN-TL descriptor are 128, 256, 96, and 224 per atom for MACE-MP0-small, MACE-MP0-large, MACE-OFF23-small, and MACE-OFF23-large, respectively.

Using GNN-TL descriptors as input, a regression model was constructed to predict NMR chemical shielding constants. For the regressor, one can choose methodologies, such as GPR, KRR, or feed-forward neural network (NNs), which are contingent on the specific task. In this study, to ensure a maximally fair comparison with other descriptor-based techniques, we adopted KRR.

KRR combines the merits of ridge regression, which offers regularization to mitigate overfitting, with the kernel method, facilitating nonlinear regression. In kernel methods, the data —in the context of our study, the GNN-TL descriptors— are mapped into a high-dimensional feature space through a non-linear kernel function. The Laplacian and Gaussian kernels were applied:
\begin{equation}
k\left(\mathbf{G}_i, \mathbf{G}_j\right)=\exp \left(-\gamma \left\|\mathbf{G}_i-\mathbf{G}_j\right\|_p^p\right),
\label{eq:1}
\end{equation}
where $\gamma$ is the hyperparameter of the kernel and \(p\) is the norm parameter that differentiates the type of kernel: \(p=1\) for the Laplacian kernel and \(p=2\) for the Gaussian kernel.
In KRR, the predicted value $\hat{\sigma}_t$ for the target chemical property of the target atom is derived from the GNN-TL descriptor $\mathbf{G}_t$ as follows:
\begin{equation}
\hat{\sigma}_t\left(\mathbf{G}_t\right)=\sum_i^N \alpha_i k\left(\mathbf{G}_i, \mathbf{G}_t\right)
\label{eq:2}
\end{equation}
Here, $\alpha_i$ represents the $i^{th}$ element of the regression coefficient vector, $\boldsymbol{\alpha}$, of size $N$. The regression coefficients are determined by solving a ridge-regularized least-squares problem, which can be reduced to:
\begin{equation}
\boldsymbol{\alpha}=(\mathbf{K}+\lambda \mathbf{I})^{-1} \boldsymbol{\sigma}
\label{eq:3}
\end{equation}
where $\mathbf{I}$ denotes the identity matrix, $\boldsymbol{\sigma}$ denotes the chemical properties of each $N$ training data samples, and $\lambda$ denotes the regularization parameter. The matrix $\mathbf{K}$, is a kernel matrix, with elements given by $k\left(\mathbf{G}_i, \mathbf{G}_j\right)$.

All computations related to the KRR were executed using Scikit-learn v.1.2.2,\cite{scikit-learn} and the hyperparameters of each model were tuned using Optuna v.2.10.\cite{akiba2019optuna} For dataset sizes of up to 50K items, we conducted hyperparameter optimization for 100 iterations with ten-fold cross-validation, while for those at 100K, we limited the optimization to 10 iterations.

The quantum-kernel method leverages quantum computers to compute kernels, \cite{Sajjan_2022, Schuld_2019, Kusumoto_2021} which is achieved by embedding feature vectors generated by classical computers into quantum states. This method calculates the inner product of these quantum states to derive the desired kernels. Embedding feature vectors into quantum states corresponds to mapping them onto a Hilbert space with dimensions raised to the power of two quantum bits (qubits). Using the kernel matrix constructed on a quantum computer, we performed a KRR, denoted as quantum KRR (QKRR). 

In this study, we adopted the natural parameterized quantum circuits (NPQC) Kernel, which has been demonstrated to possess performance characteristics similar to the Gaussian kernel, both theoretically and in actual hardware experiments \cite{Haug_2023, Haug_2022,Benedetti_2019}. All computations were conducted using Scikit-qulacs \cite{scikit-learn, qulacs_osaka, Suzuki_2021}. The quantum kernel was constructed in a 10-qubit space. Hyperparameters for the quantum kernel were determined through grid search. The determined parameters of NPQC kernel were $c$ = 1.5 and the repetition times of embedding 40. The regularization hyperparameter in QKRR was determined using 10 iterations of randomized search.

\section{\label{sec:Results}Results}
In Section ~\ref{sec:dimensional_efficiency}, because we deal with many elements, we compared the dimensional efficiency of our proposed GNN-TL descriptor to well-established physics-inspired descriptors. Note that the GNN-TL descriptor can better handle complex chemical systems by exploiting the GNN-based IAP architecture.

In Section ~\ref{sec:prediction_accuracy}, we focused on the accuracy of the GNN-TL descriptor in predicting NMR chemical shifts, which are key to understanding molecular details ($e. g.$, interatomic distances and bond angles). This scenario provides an ideal test for determining how well the GNN-TL descriptor works in our study.

Our analysis began by comparing quantum kernel learning, in which the kernels are tested using a quantum computer emulator with traditional kernel learning methods. We then checked the accuracy of the GNN-TL descriptors across the different pretrained GNN models.

Finally, we juxtaposed our GNN-TL descriptor using well-established physics-inspired descriptors. This comparison demonstrates the superiority of the proposed descriptor in terms of efficiency and accuracy. Furthermore, it highlights its potential for accurately predicting chemical properties, which is crucial for advancing research in the molecular and material sciences.

\footnotetext{\footnote[2]~SOAP and FCHL were generated by Dscribe 0.4.0\cite{HIMANEN2020106949} and QML 0.4.0.12\cite{qml}, respectively. The default hyperparameters were selected as in QM9NMR paper.}

\subsection{\label{sec:dimensional_efficiency}Dimensional Efficiency}

\begin{table*}[]
\caption{\label{tab:1}
Scaling of descriptor dimensions with respect to number of elemental species $N_{\mathrm{elem}}$} 
\begin{tabular}{ccccccccccc}
\hline
&SOAP\footnote[2]&FCHL19\footnote[2]&\begin{tabular}[c]{@{}c@{}}SchNet\\ GNN-TL\end{tabular}&\begin{tabular}[c]{@{}c@{}}MEGNet\\ GNN-TL\end{tabular}&\begin{tabular}[c]{@{}c@{}}M3GNet\\ GNN-TL\end{tabular}&\begin{tabular}[c]{@{}c@{}} MACE-MP0- \\ small GNN-TL\end{tabular}&\begin{tabular}[c]{@{}c@{}} MACE-MP0- \\ large GNN-TL\end{tabular}&\begin{tabular}[c]{@{}c@{}} MACE-OFF23-\\ small GNN-TL\end{tabular}&\begin{tabular}[c]{@{}c@{}}  MACE-OFF23-\\ large GNN-TL\end{tabular}\\
\hline
$N_{\mathrm{elem}}$&$O(N^{2}_{\mathrm{elem}})$&$O(N^{2}_{\mathrm{elem}})$&$O(1)$&$O(1)$&$O(1)$&$O(1)$&$O(1)$&$O(1)$&$O(1)$\\
5&5,740&740&128&32&64&128&256&96&224\\
10&22,680&2,440&-&-&64&128&256&96&224\\
89&1,737,120&162,336&-&-&64&128&256&-&-\\
\hline
\end{tabular}
\end{table*}

At the atomic level, descriptors are tools designed to encode information about atoms within molecules or crystalline materials into vectors. Popular descriptors, such as SOAP and FCHL18, excel at intricately capturing the environment within an atom's cutoff radius. Although these descriptors have achieved significant success in various accuracy benchmarks, they also present challenges due to their large dimensions. Various strategies have been developed to address these challenges,\cite{lopanitsyna2023modeling, willatt2018feature, li2022encoding, Christensen_2020} including refining the descriptor itself, using principal component analysis for dimensionality reduction, and exploring NNs to encode them. In particular, Christensen et al. applied Behler's method of the atom-centered symmetry function \cite{Behler_2011} for NN potential to discretize FCHL18\cite{Faber_2018} to derive a compact and accurate FCHL19.\cite{Christensen_2020}

In Table 1, we present the scaling of the SOAP, FCHL19 and various GNN-TL descriptors in response to an increase in the number of elemental species considered. Additionally, for the QM9, QMugs,\cite{isert2022qmugs} and MPF.2021.8 or MPtrj datasets,\cite{Chen_2022} the descriptor dimensions corresponding to 5, 10, and 89 elemental species comprising each dataset are summarized, respectively. Remarkably, with an increase in the number of element types, both SOAP and FCHL19 exhibited quadratic scaling. As a snapshot, when representing five elements in the QM9 dataset, the SOAP and FCHL19 methods have dimensions of 5,740 and 740, respectively. This dimensional disparity increases with the number of elemental types. Hence, to represent the 89 elements, the dimensions increased to 1,737,120 and 162,336, respectively. These dimensions are hundreds to tens of thousands of times larger than the compact GNN-TL descriptors, which ranges from 64 to 256 dimensions. Owing to its consistent dimensionality, irrespective of the increase in elements, the GNN-TL descriptors are overwhelmingly compact.

\subsection{\label{sec:prediction_accuracy} Prediction Accuracy: NMR Chemical Shifts}
The NMR chemical shifts, $\delta$, were predicted using the chemical shielding constant of the reference substance, $\sigma_\mathrm{ref}$, as the baseline. The NMR chemical shift was calculated using the following equation: 
\begin{equation}
\delta = \sigma_\mathrm{ref} - \sigma.
\label{eq:4}
\end{equation}

The reference substances selected for the various nuclei in this study are widely recognized and commonly adopted in the literature.\cite{Xin_2017, Puzzarini_2009, Wasylishen_2002, rosenau2018exposing, Gupta_2021} Specifically, tetramethylsilane was selected for both $^1$H and $^{13}$C, nitromethane (\ce{MeNO2}) for $^{15}$N, water-$^{17}$O (H$_2$$^{17}$O) for $^{17}$O, and trichlorofluoromethane (\ce{CFCl3}) for $^{19}$F. We determined the chemical shielding constants for these well-established reference substances as follows: \SI{31.7608}{ppm} for $^1$H, \SI{187.0521}{ppm} for $^{13}$C, \SI{-147.8164}{ppm} for $^{15}$N, \SI{325.8642}{ppm} for $^{17}$O, and \SI{171.2621}{ppm} for $^{19}$F. These constants were evaluated by calculations at the mPW1PW91\cite{adamo1998exchange}/6-311+G(2d,p) level using density functional theory (DFT) and gauge-including atomic orbital (GIAO)\cite{ditchfield1972molecular} methods. Structure optimization was conducted at the B3LYP\cite{stephens1994ab}/6-31G(2df,p) level in alignment with the methodologies employed for the QM9 NMR dataset. All calculations were performed using the Gaussian 16 software suite.\cite{g16}

In our study, we utilized the QM9NMR dataset, which contains approximately 134K small organic molecules containing C, N, O, and F (excluding H), with each molecule having no more than nine atoms.\cite{Gupta_2021, Ramakrishnan_2014} This dataset provides the detailed NMR chemical shielding constants for these molecules. To analyze how the model accuracy changes with training data size, we adopted an approach similar to that used in the original publication of the QM9NMR dataset.\cite{Gupta_2021} Specifically, for $^{13}$C, of a total of 831K data points, we randomly withheld 50K data points to build our test set. Subsequently, from the remaining $^{13}$C NMR chemical shifts, we randomly selected subsets containing 100, 200, 500, 1K, 2K, 5K, 10K, 50K, 100K, and 200K data points to create various training sets. For the other isotopes ($i.e.$, $^{1}$H, $^{15}$N, $^{17}$O, and $^{19}$F), the test sets were similarly established by withholding 50K, 30K, 50K, and 1K data points, respectively. The training size for $^{19}$F was set to 2K, whereas the other isotopes were trained on datasets of 100K data points. In addition to the QM9 NMR dataset, we sought to validate the performance of our model on external datasets. Hence, we employed the two sets of molecules provided in another study;\cite{Gupta_2021} one consisting of 40 drug molecules from the GDB17 universe and another containing 12 drugs with 17 or more heavy atoms. 

\begin{figure}[h]
\includegraphics{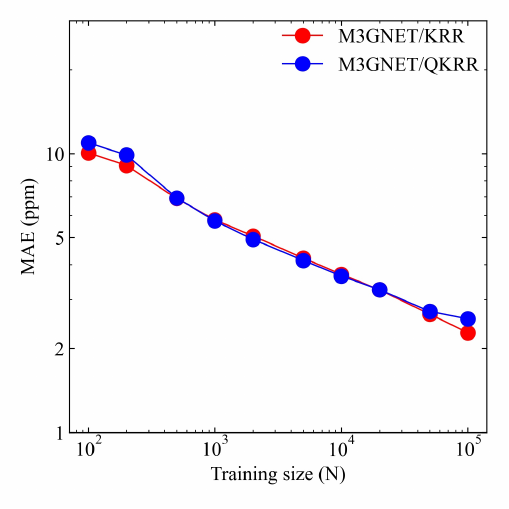}
\caption{\label{fig:2}Log-log plot of the training size ($N$) and MAE for the $^{13}$C NMR chemical shielding constant prediction model. The red and blue colors represent the results of the KRR with the Laplacian kernel and QKRR with the NPQC kernel using GNN-TL descriptors from the pre-trained M3GNet model, respectively.}
\end{figure}

Fig. 2 shows the relationship between the mean absolute error (MAE) for the $^{13}$C NMR shielding constant predictions and the training data size. Both QKRR and KRR demonstrated consistent improvements in predictive accuracy with an increase in training size. Notably, the quantum kernel exhibited a performance comparable to that of the Laplacian kernel. For a training size of 100K, the MAE for the $^{13}$C predictions was \SI{2.28}{ppm}. In a comparative study by Gupta et al., the KRR models using the Coulomb matrix (CM),\cite{Rupp_2012} SOAP, and FCHL descriptors reported MAEs of approximately 4, 2.1, and \SI{1.88}{ppm}, respectively, for the same training size.\cite{Gupta_2021} Compared with the CM descriptor, our GNN-TL descriptor showed significantly better predictive capabilities, achieving an MAE that was nearly half that of the CM descriptor. Although our method did not exceed the accuracy levels of SOAP and FCHL, the performance of the GNN-TL descriptor was competitive, highlighting its potential as a robust descriptor.

\footnotetext{\footnote[3]~The value is taken from \cite{El_Samman_2023}}

\begin{table}[b]
\centering
\caption{\label{tab:2}
The architecture dependence of the predictive performance. For KRR, the Gaussian kernel was applied.}
\begin{tabular}{cc}
\hline
GNN-TL descriptor/Regressor & RMSE (ppm)\\

\hline
SchNet/NN\footnote[3]&12.8\\
MEGNet/KRR&20.08±0.55\\
M3GNet/KRR&10.02±0.37\\
MACE-MP-0-small/KRR&9.77±0.34\\
MACE-MP-0-large/KRR&9.74±0.27\\
MACE-OFF23-small/KRR&8.05±0.19\\
MACE-OFF23-large/KRR&8.15±0.42\\

\hline
\end{tabular}
\end{table}

Next, we compared the performance of the GNN-TL descriptors derived from different IAP architectures. Recently, independent of our work, a predictive model for $^{13}$C NMR chemical shielding was proposed using a pretrained IAP known as SchNet, which is a pioneering GNN used as a descriptor.\cite{El_Samman_2023} This model was trained on 400 data points of $^{13}$C NMR chemical shielding constants of the molecules in QM9 dataset,\cite{Ramakrishnan_2014} with the SchNet GNN-TL descriptor as an input to a feed-forward NN for regression. The predictive accuracy of the SchNet/NN was a root mean-squared error (RMSE) of \SI{12.8}{ppm}. In pursuit of a fair comparison with their model, we applied KRR using pre-trained MEGNet, M3GNet and MACE GNN-TL descriptors, setting our training data size to 400 data points of $^{13}$C NMR chemical shielding constants. To account for the influence of random sampling, we created 10 different training sets, each comprising 400 data points. The effect of potential data bias was then quantified by calculating the mean RMSE and standard deviation (STD) for each model. Detailed verification including kernel function dependencies can be found in the Appendix. The results of this comparative study are summarized in Table 2. In Table 2, the results for KRR using the Gaussian kernel, which showed superior accuracy compared to the Laplacian kernel, are presented.

In contrast to the SchNet/NN model's RMSE of 12.8 ppm, the MEGNet/KRR model shows significantly lower predictive accuracy with an RMSE of 20.08 ± 0.55 ppm, suggesting that the MEGNet descriptor is less effective for $^{13}$C NMR chemical shielding data. The M3GNet/KRR model demonstrates a substantial improvement with an RMSE of 10.02 ± 0.37 ppm. Models using MACE descriptors show even greater accuracy: the MACE-MP-0-small/KRR and MACE-MP-0-large/KRR models achieve RMSEs of 9.77 ± 0.34 ppm and 9.74 ± 0.27 ppm, respectively. The best performance is observed with the MACE-OFF23-small/KRR model, which has an RMSE of 8.05 ± 0.19 ppm, with the MACE-OFF23-large/KRR model close behind at 8.15 ± 0.42 ppm. These results highlight the superior performance of the MACE descriptors, particularly MACE-OFF23-small, in enhancing the accuracy of KRR models for predicting $^{13}$C NMR chemical shielding. A more detailed discussion of the nuances of these architectural differences is presented in Section ~\ref{sec:influence_of_architecture}.

The accuracy of KRR models incorporating the M3GNet GNN-TL descriptor with a Laplacian kernel for NMR chemical shifts was evaluated for each test set of the five different nuclei. Table 3 lists the statistical performance metrics for predicting NMR chemical shifts. Across all elements, the MAE for the test set remained below \SI{5}{ppm}. The MAE for $^1$H and $^{19}$F were notably low at \SI{0.18}{ppm} and \SI{2.65}{ppm}, respectively, indicating a high degree of prediction accuracy for these nuclei in the unseen molecular environments. The MAE for $^{17}$O, although higher at \SI{4.95}{ppm}, still reflects a reasonable predictive capability, given the complexity of the oxygen chemical shifts. The STD and interquartile range (IQR) values in the Table 3 represent the distribution of chemical shifts within the training data, rather than the accuracy of the model itself. Thus, the higher STD and IQR values for $^{17}$O do not indicate a lack of model precision but rather the natural variability inherent in the $^{17}$O chemical shifts within the training data. The MAE/STD ratio can still offer insights into model performance relative to data variability. For example, the relatively low ratio of $^{17}$O (2.21\%) suggests that the model predictions are consistent with the diversity of the training data. On the other hand, the higher ratios for $^1$H (9.09\%) and $^{19}$F (7.78\%) indicate that the accuracy of the models are not as high as desired, particularly when considering the range of chemical shifts represented in the training dataset. The maximum absolute error (MaxAE) for all nuclei is comparable to the STD of the training data. This is attributed to random sampling and is expected to improve with the application of more sophisticated data point sampling techniques, such as active learning.

Subsequently, these models were employed to predict the NMR chemical shifts of a single molecule  \( \mathrm{C_5H_5N_2OF} \) containing five elements that was not included in the training data. The results are shown in Fig. 3. The MAE for each nucleus were found to be \SI{0.08}{ppm} for $^1$H, \SI{1.03}{ppm} for $^{13}$C, \SI{6.45}{ppm} for $^{15}$N, \SI{2.86}{ppm} for $^{17}$O, and \SI{6.73}{ppm} for $^{19}$F. The remarkably low MAE for $^1$H and $^{13}$C underscores the high accuracy of our model for these nuclei, with predictions that closely mirror the calculated values. The model performed well for the more challenging $^{15}$N and $^{17}$O nuclei, where the chemical shifts can be significantly affected by subtle changes in the molecular structure and environment, as indicated by the MAE values. The $^{19}$F nucleus, while having a higher MAE, showed excellent agreement with the DFT/GIAO calculations, suggesting that the model predictions were robust, even for nuclei with typically higher chemical shift ranges. These results demonstrate the strong predictive power and potential of the model as a reliable tool for accurately predicting NMR chemical shifts across a variety of nuclei, even in molecules beyond the scope of the training data. 

\begin{table}[]
\centering
\caption{\label{tab:3}
Predictive performance and data variability of NMR shielding constants for 5 elements}
\begin{tabular}{cccccc}
\hline
&$^1$H&$^{13}$C&$^{15}$N&$^{17}$O&$^{19}$F\\

\hline
MAE (ppm)&0.18&2.28&3.42&4.95&2.65\\
MaxAE (ppm) &7.50&68.58&71.62&279.84&39.31\\

STD (ppm)&1.98&51.96&119.58&224.40&34.07\\
IQR (ppm)&2.34&59.93&211.19&354.25&36.77\\
MAE/STD ($\%$) &9.09&4.38&2.86&2.21&7.78\\
\hline
\end{tabular}
\end{table}

\begin{figure}[]
\centering
\includegraphics{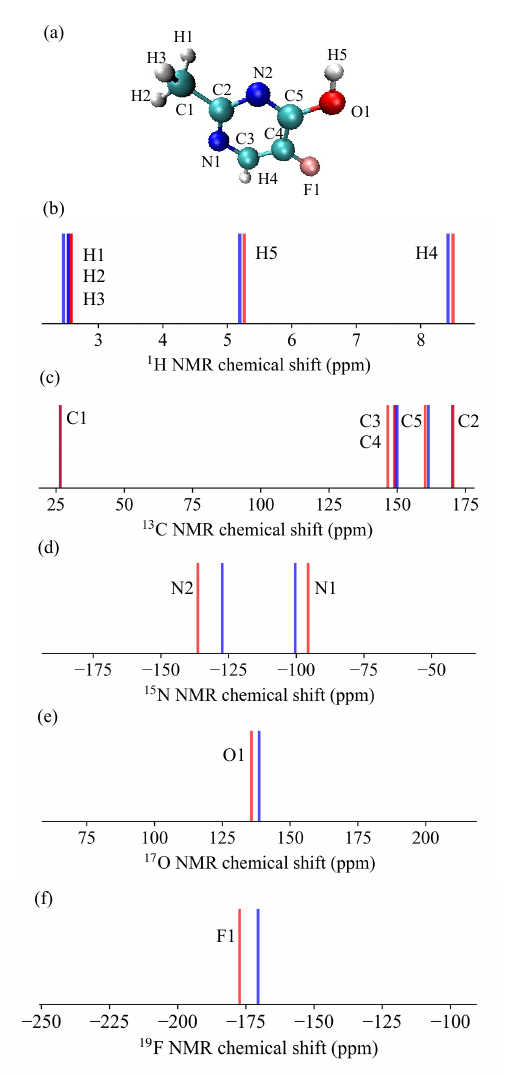}
\caption{\label{fig:Fig3}Predicted NMR chemical shifts for (a) a single molecule, randomly selected from the QM9NMR dataset and not included in the training data, for (b) $^1$H, (c) $^{13}$C, (d) $^{15}$N, (e) $^{17}$O, and (f) $^{19}$F. These predictions (represented by red lines) are compared with the calculated values at the DFT/GIAO level, which are considered as the correct values (depicted by blue lines).
}
\end{figure}

\begin{table}[t]
\centering
\caption{\label{tab:4}
The MAE values for the prediction of the 50K QM9NMR hold out set, 40 drug molecules from GDB17 Universe and the other containing 12 drugs with 17 or more heavy atoms. The values in parentheses indicate MaxAE. All units are in ppm.}
\begin{tabular}{cccc}
\hline
&FCHL\footnote[4]&\begin{tabular}[c]{@{}c@{}}M3GNet\\ GNN-TL\end{tabular} &\begin{tabular}[c]{@{}c@{}}MACE-OFF23-small\\ GNN-TL\end{tabular} \\
\hline 
50K QM9&1.88 &2.28 (68.58)&1.87 (59.76) \\
40 drugs&3.7&3.46 (29.86)&2.83 (16.08)\\
12 drugs&4.2&4.21 (20.48)&3.85 (24.70)\\
\hline
\end{tabular}
\end{table}

We then expanded our assessment to evaluate the predictive ability of our model for molecules larger than those in the QM9 NMR dataset. As such, we incorporate the test sets provided in Ref. \cite{Gupta_2021}, which comprised 40 drug molecules from the GDB17 universe and another set containing 12 drugs with 17 or more heavy atoms. See Ref. \cite{Gupta_2021} for the structures of these molecules.

Table 4 presents the benchmark results for each test set using our M3GNet GNN-TL descriptor and MACE-OFF23-small GNN-TL descriptor. For comparison, we used the FCHL descriptor from Gupta’s study.\cite{Gupta_2021} To ensure a fair comparison, we employed our GNN-TL descriptor models trained on a size of 100K $^{13}$C chemical shielding constants. For both models, an increased molecular size in the dataset correlated with deterioration of the MAE value. Notably, although our M3GNet GNN-TL descriptor did not match the \SI{1.88}{ppm} value achieved by the FCHL descriptor for the QM9 50K test set, our model exhibited an MAE value that was approximately \SI{0.3}{ppm} lower for the 40 GDB17 dataset test. The MACE-OFF23-small GNN-TL descriptor showed even better performance, with an MAE of \SI{1.87}{ppm} for the QM9 50K test set, closely matching the FCHL descriptor, and significantly outperforming it for the 40 GDB17 dataset with an MAE of \SI{2.83}{ppm}. For the set of 12 drugs with 17 or more heavy atoms, the M3GNet descriptor showed an MAE of \SI{4.21}{ppm}, while the MACE-OFF23-small descriptor showed an MAE of \SI{3.85}{ppm}. Notably, the M3GNet descriptor's accuracy is comparable to the FCHL descriptor. The results were nearly identical for the set of 12 drugs with 17 or more heavy atoms, highlighting that the M3GNet GNN-TL descriptor was less affected by increasing molecular size. On the other hand, the MACE-OFF23-small descriptor significantly outperforms FCHL with an MAE of \SI{3.85}{ppm}, highlighting its superior predictive performance.

For a detailed comparison, Fig. 6 illustrates the molecule-specific MAE values for both drug test sets. The molecular structures are provided in Ref.\cite{Gupta_2021}. Our M3GNet and MACE-OFF-small GNN-TL descriptor-based prediction models ensured that the highest MAE values for individual molecules across both test sets remained below \SI{10}{ppm}. Intriguingly, the desflurane molecule, which posed the greatest challenge, showed MAE values of \SI{53.3}{ppm}, \SI{9.35}{ppm} and \SI{8.31}{ppm} for the FCHL, M3GNet and MACE-OFF23-small GNN-TL descriptor models, respectively. This suggests an approximately 80\% reduction in the MAE with our descriptor, which is likely attributable to differences in the encompassed descriptor domain.

\footnotetext{\footnote[4]~FCHL results are taken from \cite{Gupta_2021}.}

The cutoff radius for the FCHL descriptor was determined through a grid search,\cite{Gupta_2021} which settled at \SI{4.0}{\angstrom}. In this scenario, the two fluorine atoms in the terminal trifluoromethyl group (\ce{CF3}) of the desflurane molecule, which lie beyond \SI{4}{\angstrom} from the \ce{CF2H} carbon, were neglected. In contrast, our M3GNet descriptor had a \SI{6}{\angstrom} cutoff radius during the initial graph configuration and a \SI{5}{\angstrom} cutoff for three-body interactions during graph convolution, capturing the entire \ce{CF3} group. This suggests that the descriptor adequately accounts for the influence of the terminal trifluoromethyl group. Additionally, the intrinsic ability of GNN-TL descriptors to account for environments beyond their cutoff radius, owing to graph convolution, may have contributed to the substantial improvement in MAE. Notably, the MACE-OFF23-small model, with a cutoff value of \SI{4.5}{\angstrom}, achieves the highest accuracy, even though it does not capture the fluorine element at a distance of \SI{4.65}{\angstrom} in the \ce{CF3} group. In summary, the proposed M3GNet and MACE GNN-TL descriptors demonstrate the capability of predicting $^{13}$C NMR chemical shifts for molecules outside the training dataset with an accuracy comparable to that of the state-of-the-art FCHL descriptor. 

Lastly, to explore further practical applications of the constructed models, we validated the NMR chemical shielding constants obtained using semi-empirical PM7-level geometries as inputs against the NMR chemical shift values obtained using DFT/GIAO-level structures from the training data. This validation was performed on the QM9 50K holdout set and two drug molecule test sets, as provided by Ref.\cite{Gupta_2021}. The $^{13}$C prediction model employed was the M3GNet/KRR model. The MAE values for each molecule in the drug datasets can be found in Figures 6(b) and 6(d). For the QM9 50K holdout set, the result was \SI{3.61}{ppm}, showing a significant deterioration of \SI{1.33}{ppm} compared to when DFT-level geometries were used as inputs. Conversely, predictions for the 40 drugs and 12 drugs test sets showed only minor deteriorations of \SI{0.23}{ppm} and \SI{0.04}{ppm}, respectively. These results suggest that even when using more readily available PM7-level geometries as inputs, the transferability of the model remains robust for extrapolative predictions on larger molecules compared to the training data.

\section{\label{sec:Discussion}Discussion}

\subsection{\label{sec:influence_of_architecture}Influence of Architectural Choices on GNN-TL Descriptor Performance}

In our exploration of different architectures for generating GNN-TL descriptors, we observed several patterns. First, as shown in Table 1 and Table 2, it is important to note that the accuracy of GNN-TL descriptors does not necessarily improve with an increase in the dimensionality of the descriptors. With this in mind, we discuss the architecture of each GNN-based IAP. SchNet, which operates on GNN-based local descriptors to evaluate systems as summations of atomic energies, accounts only for pairwise interactions. This limited inclusion could potentially constrain expressions, leading to inadequate representational power. The subpar performance of MEGNet during transfer learning may be attributed to its architectural design as it integrates atomic (local) descriptors into molecular (global) descriptors through concatenation. This means that the final piece of information passed to the MLP is not extracted directly from the end of the model, which might not be the optimal representation for targeted atomic-wise property prediction; however, it is expected to be suitable for molecule-wise property predictions. Moreover, the M3GNet architecture, which considers three-body interactions, has the potential to capture the three-dimensional structure of molecules with high resolution. Additionally, the MACE model, an E(3) equivariant GNN, has demonstrated high performance as an IAP, suggesting that the outputs of its GNN layers are highly accurate in representing molecular structures. Furthermore, future improvements in accuracy may be achieved by leveraging the outputs of higher-order GNN layers in the MACE model, corresponding to the two-body and three-body terms in the atomic cluster expansion.

\subsection{\label{sec:siginificance_of_dataset}Significance of Dataset Size and Diversity}

The M3GNet training regimen incorporates data from 187,687 ionic steps spanning 62,783 compounds, including 187,687 energies, 16,875,138 force components, and 1,689,183 stress components. This diverse dataset covers 89 elements from the periodic table. The model is not limited to learning only the energies associated with these elements but extends to atomic-level forces. Moreover, M3GNet training includes not only stable structures but also the processes of structure optimization. The ingestion of vast amounts of data from crystalline systems may have endowed the M3GNet with enhanced expression, potentially making it adept at interpolating molecular systems. The pre-trained MACE-MP0 model was trained using ten times more energy data of crystalline systems, potentially contributing to the improved accuracy of the $^{13}$C NMR chemical shift predictions shown in Table 2. On the other hand, the MACE-OFF23 model, which is specialized for molecules containing 10 elemental species, was trained on a dataset comprising about 1M energy data points, with structures containing up to 150 atoms. This extensive training dataset might make it more suitable for predicting molecular NMR chemical shifts. Thus, the training data for IAPs, much like their architectures, could be a crucial factor in determining the performance of the descriptors.

\subsection{\label{sec:potential_for_transfer}Potential for Transfer Learning on Quantum Computer}

There is a potential for leveraging quantum computation approaches.\cite{Cerezo_2021} Specifically, our 10 qubit QKRR, facilitated by a simulator, demonstrated a performance comparable to that of state-of-the-art KRR. This is underpinned by the theoretical equivalence of the NPQC with the Gaussian kernel. The quantum kernel method stands out because of its capability to compute with fewer measurement iterations than other quantum computation methodologies, such as quantum neural networks.\cite{Mitarai_2018} In particular, our proposed M3GNet GNN-TL descriptor can be feasibly realized with a minimum of six qubits, enabling evaluations with a quantum bit count that is more efficient than traditional descriptors, such as SOAP. However, embedding for higher-dimensional SOAP appears to be a challenge, possibly due to noise. From a futuristic perspective, there is excitement about the possibility of developing kernels that traditional computers cannot express, as well as accelerating the inversion calculations of kernel matrices using quantum algorithms. The constant scaling property of our proposed method concerning element number dimensions may significantly contribute to real-time material exploration powered by quantum-classical hybrid algorithms in the near future.

\section{\label{sec:Conclusion}Conclusion}

The dynamics of machine learning and its extensive applications across various domains are driving cutting-edge research. Our endeavor to integrate transfer learning with pretrained IAP GNNs for NMR chemical shift prediction offers a paradigm shift in efficiency and scalability. The GNN-TL descriptor presents an unparalleled advantage in terms of scalability due to its consistent dimensionality, irrespective of the number of elements.

Comparative evaluations with other renowned descriptors, such as SOAP, suggest that the GNN-TL descriptor can match, if not surpass, the performance of its contemporaries while maintaining a more compact representation. This is especially important when factoring large datasets, where dimensionality can exponentially burgeon.

Architectural choice plays a pivotal role in the performance of GNN-TL descriptors. Moreover, the diversity and vastness of the training dataset, which encompasses myriad elemental types and structural configurations, augment the robustness and versatility of the GNN. 

Our proposed model has immense potential for creating a unified framework capable of predicting various atomic and molecular properties simultaneously, presenting profound implications for accelerated material and molecular research. This potential union of multiple predictions can usher in an era of comprehensive understanding and quicker innovations, possibly revolutionizing fields, such as catalysis, drug discovery, and material design.

The union of transfer learning with pretrained GNNs not only augments prediction accuracy but also drastically reduces learning costs, presenting a cost-effective and efficient alternative to more computationally intensive methods. As we move toward an era in which data-driven insights and models govern the pace of innovation, our research offers a promising pathway for future endeavors in the domain of chemical property predictions with both classical and quantum computers.

\textit{Note added -} As we were finalizing this manuscript, we became aware of recent articles \cite{El_Samman_2023, El_Samman_2023_2, elijovsius2024zero} that also utilize intermediate information from graph neural network potentials. 
In Section ~\ref{sec:prediction_accuracy}, we added a direct comparison between our results and theirs. 
Elijo\v{s}ius et al. applied the pre-trained MACE descriptor to generative modeling of molecules \cite{elijovsius2024zero}.

\section*{Appendix}

\subsection*{Distribution of Datasets for Each NMR Chemical Shift Prediction Model}

The distributions of the training and test sets sampled from the  QM9NMR dataset are shown in Fig. 4. Fig. 4(a) shows that above 5K, the distribution is in good agreement with the overall distribution of the $^{13}$C NMR shielding constants. For the other elemental species, the distributions of the training and test sets were in good agreement with the overall distribution.

\subsection*{\label{sec:kernel_dependency}Kernel Function Dependency for Various GNN-TL Descriptors}

\begin{table}[b]
\centering
\caption{\label{tab:5}
Accuracy (measured by RMSE) of GNN-TL/KRR models trained on 400 13C NMR chemical shift values for different kernel functions. All units are in ppm.}
\begin{tabular}{ccc}
\hline
GNN-TL descriptor& Gaussian kernel & Laplacian
 kernel\\

\hline
MEGNet&{\textbf{20.08±0.55}}&21.12±0.56\\
M3GNet&\textbf{10.02±0.37}&10.31±0.38\\
MACE-MP-0-small&\textbf{9.77±0.34}&10.78±0.31\\
MACE-MP-0-large&\textbf{9.74±0.27}&10.17±0.30\\
MACE-OFF23-small&\textbf{8.05±0.19}&8.64±0.13\\
MACE-OFF23-large&\textbf{8.15±0.42}&8.77±0.21\\

\hline
\end{tabular}
\end{table}

The accuracy of KRR models using Gaussian and Laplacian kernels was evaluated. Table 5 presents the mean RMSE and its standard deviation for predictions on the 50K holdout set by models trained on 400 data points of ¹³C using Gaussian and Laplacian kernels. For all models using GNN-TL descriptors, the mean RMSE of models with Gaussian kernel was found to be more accurate than those with Laplacian kernel. However, the variation in accuracy due to dataset sampling (standard deviation) was found to have a greater impact than kernel choice in models with MEGNet and M3GNet GNN-TL descriptors. On the other hand, in models with MACE GNN-TL descriptors, the impact of kernel choice was more significant than the variation due to dataset sampling.

Next, Table 6 shows the accuracy of KRR models using M3GNet and MACE-OFF23-small GNN-TL descriptors trained on a 100K ¹³C training set. Unlike models trained on the 400 ¹³C training set, the KRR models with M3GNet GNN-TL descriptors consistently showed higher accuracy with the Laplacian kernel compared to the Gaussian kernel. Conversely, the results for MACE-OFF23-small GNN-TL descriptors were similar to those for models trained on the 400 ¹³C training set, with the Gaussian kernel models demonstrating higher accuracy. This suggests that the appropriate kernel function may vary depending on the size of the training data.

Finally, these results indicate the choice of kernel functions for KRR models as presented in the Results section of this paper. For models trained on 400 ¹³C data points, all KRR models using GNN-TL descriptors employed the Gaussian kernel. In contrast, for models trained on 100K ¹³C data points, the Laplacian kernel was used for KRR models with M3GNet GNN-TL descriptors, whereas the Gaussian kernel was employed for models with MACE-OFF23-small GNN-TL descriptors.

\begin{table}[]
\centering
\caption{\label{tab:6}
The kernel function dependency of accuracy (MAE) for the prediction of the 50K QM9NMR hold out set, 40 drug molecules from GDB17 Universe and the other containing 12 drugs with 17 or more heavy atoms. All units are in ppm.} 
\begin{tabular}{ccccc}
\hline
&\multicolumn{2}{c}{M3GNet}&\multicolumn{2}{c}{MACE-OFF23-small} \\
&Gaussian&Laplacian&Gaussian&Laplacian\\
\hline 
50K QM9&2.35&\textbf{2.28}&\textbf{1.87}&2.10 \\
40 drugs&3.98&\textbf{3.46}&\textbf{2.83}&3.21 \\
12 drugs&5.14&\textbf{4.21}&\textbf{3.85}&3.93 \\
\hline
\end{tabular}
\end{table}

\subsection*{Validation of Learning Accuracy of NMR Chemical Shift Prediction}

Fig. 5 illustrates the accuracy of the KRR models trained using the M3GNet GNN-TL descriptor for five elemental species. The MAE values for the NMR shielding constant, for train/test, are as follows: for $^{1}$H, 0.0344/0.1767; $^{13}$C, 0.1420/2.2798 ppm; $^{15}$N, 0.3910/3.4157 ppm; $^{17}$O 0.8881/4.9509 ppm; and $^{19}$F 0.0864/2.6518 ppm. 

\begin{figure*}[h]
\centering
\includegraphics{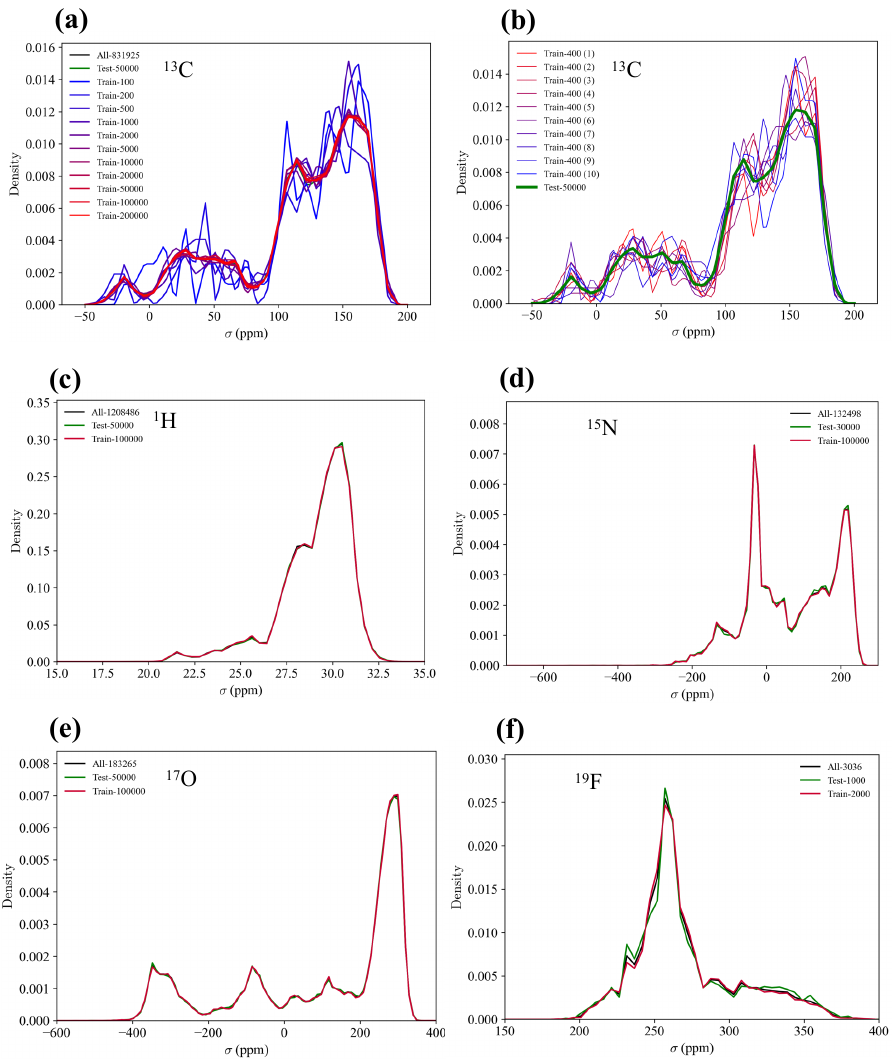}
\caption{\label{fig:4} Distributions of the NMR shielding constants of the training subsets and test set sampled from the of the QM9NMR dataset for the five elemental species: (a) $^{13}$C (for dataset size dependency), (b) $^{13}$C (for potential data bias), (c) $^1$H, (d) $^{15}$N, (e) $^{17}$O, and (f) $^{19}$F, respectively.}
\end{figure*}

The accuracy of the GNN-TL descriptors was also validated using the molecular structures of two drug molecule data sets reported in Ref 29. The predicted $^{13}$C NMR shielding constants for each drug molecule using the M3GNet and MACE-OFF23 GNN-TL/KRR models are shown in Fig. 6 (a) and (c). These predictions are accompanied by the values predicted by the FCHL/KRR model.\cite{Gupta_2021} The prediction results of the M3GNet/KRR model using PM7-level optimized geometries, along with the prediction results using DFT-level geometries, are shown in Fig. 6(b) and (d).

\begin{figure*}[h]
\centering
\includegraphics{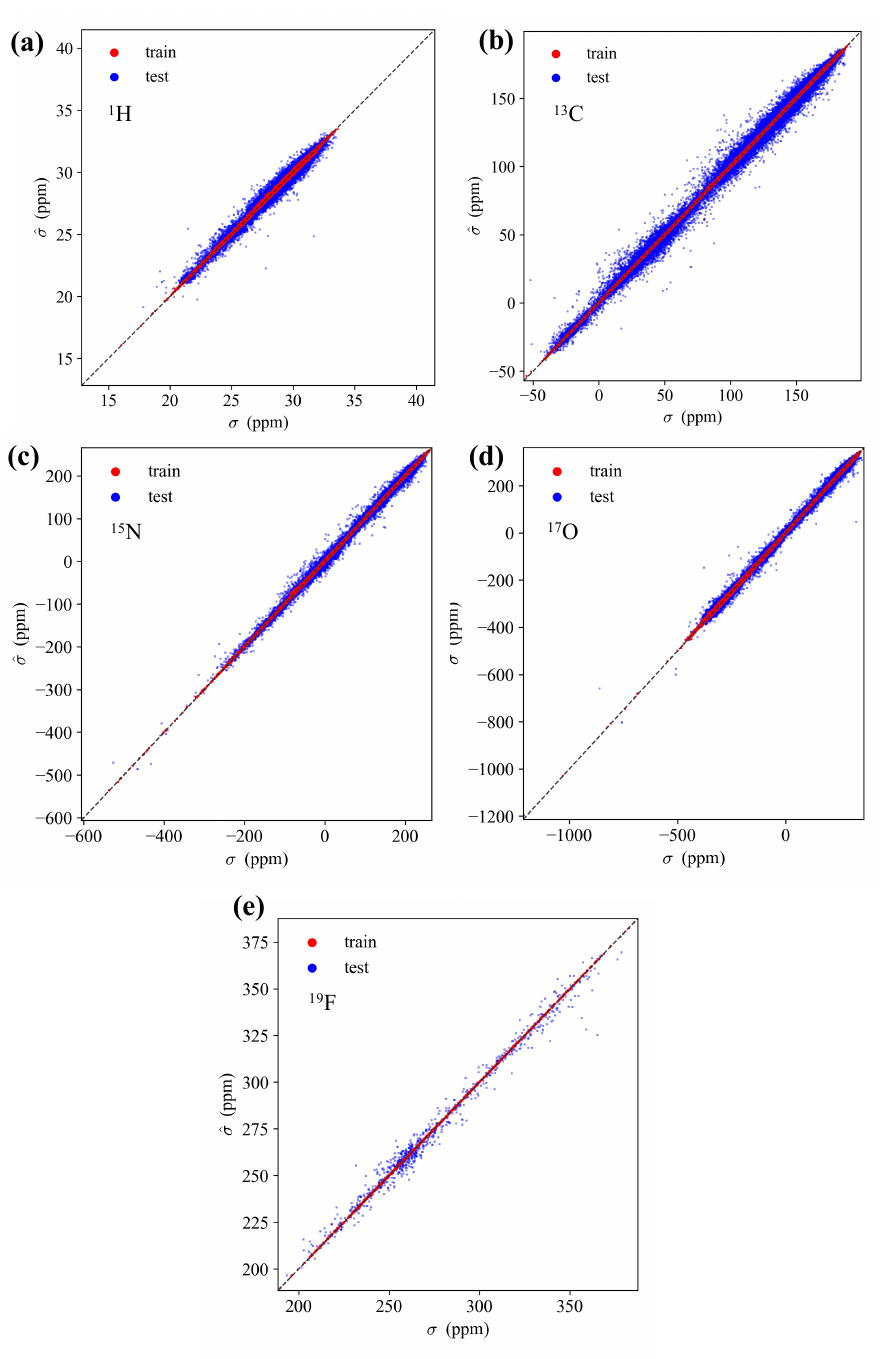}
\caption{\label{fig:5}Scatterplots for the training set (red) and test set (blue) showing NMR chemical shifts from the QM9NMR dataset, using the M3GNet GNN-TL/KRR model constructed with QM9NMR data for the five elemental species: (a) $^1$H, (b) $^{13}$C, (c) $^{15}$N, (d) $^{17}$O, and (e) $^{19}$F, respectively.}

\end{figure*}

\begin{figure*}[h]
\centering
\includegraphics{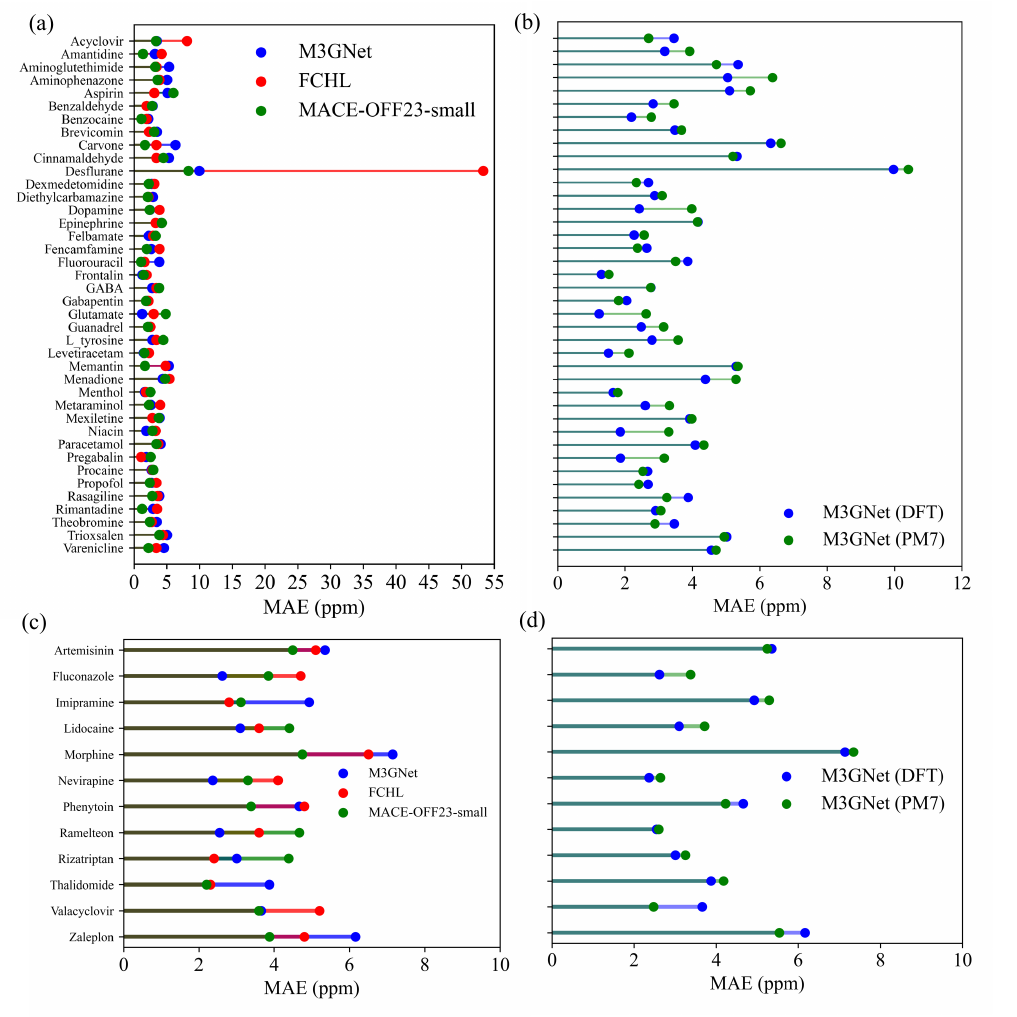}
\caption{\label{fig:6}Comparison of $^{13}$C NMR shielding constant predictions using different descriptors for (a) 40 drug molecules from the GDB17 universe and (d) 12 drugs with 17 or more heavy atoms. The predictions were made using the KRR model with the FCHL descriptor (red), the M3GNet GNN-TL descriptor (blue), and the MACE-OFF23-small GNN-TL descriptor (green). The FCHL results were taken from Ref. \cite{Gupta_2021}. The results for the M3GNet/KRR model using DFT-level geometries and PM7-level geometries are shown in (b) and (d), respectively.} 
\end{figure*}

\section*{Data Availability}

Data and code required to reproduce the figures and tables related to the GNN-TL descriptors and the NMR shielding constants prediction models presented in the manuscript is publicly accessible on GitHub at \url{https://github.com/TShiotaSS/gnn-tl}. The dataset utilized for the prediction of NMR chemical shifts, specifically the QM9NMR dataset, is available at DOI: \url{https://doi.org/10.17172/NOMAD/2021.10.16-1} and GitHub at \url{https://moldis-group.github.io/qm9nmr/}. Results of the DFT/GIAO calculations for isolated atoms, used for NMR chemical shift computations, are included within the manuscript. The GNN-TL descriptor vectors for the QM9NMR datasets are available at DOI: \url{https://doi.org/10.6084/m9.figshare.25484068.v2}. We have modified the code to extract GNN-TL descriptors from the pretrained M3GNet model on Github at https://github.com/materialsvirtuallab/m3gnet, and this adapted version can be found at \url{https://github.com/TShiotaSS/gnn-tl/tree/main/scripts/m3gnet}. The code used to extract GNN-TL descriptors from the pretrained MEGNet model can be found on GitHub at \url{https://github.com/materialsvirtuallab/megnet/blob/master/megnet/utils/descriptor.py}. The code used to generate descriptors from the pretrained MACE models can be found on GitHub at \url{https://github.com/ACEsuit/mace/blob/main/mace/calculators/mace.py}. The implementation for quantum kernel ridge regression used in this study is available at \url{https://github.com/Qulacs-Osaka/scikit-qulacs/tree/main/skqulacs/qkrr}. 

\section*{Conflicts of interest}
There are no conflicts to declare.

\section*{Acknowledgements}
We thank Nobuki Inoue and Tuan Minh Do for fruitful discussions. This project was supported by funding from the MEXT Quantum Leap Flagship Program (MEXTQLEAP) through Grant No. JPMXS0120319794, and the JST COI-NEXT Program through Grant No. JPMJPF2014. The completion of this research was partially facilitated by the JSPS Grants-in-Aid for Scientific Research (KAKENHI), specifically Grant Nos. JP23H03819 and JP21K18933.
We thank the Supercomputer Center, the Institute for Solid State Physics, the University of Tokyo for the use of the facilities. 
This work was also achieved through the use of SQUID at the Cybermedia Center, Osaka University.
%%%END OF MAIN TEXT%%%


\balance


%%%REFERENCES%%%
\bibliography{rsc} %You need to replace "rsc" on this line with the name of your .bib file
\bibliographystyle{rsc} %the RSC's .bst file

\end{document}